\documentclass[epj,final,nopacs]{svjour}
\usepackage{amsmath}
\usepackage{amsfonts}
\usepackage{amssymb}
\usepackage{graphicx}
\usepackage{dcolumn}
\usepackage[T1]{fontenc}
\usepackage{cite}

\begin{document}

\title{Study of hydrogen confined in onion shells}

\author{A. L. Frapiccini\inst{1}\thanks{\emph{e-mail: afrapic@uns.edu.ar}}\and D. M. Mitnik\inst{2}}
\institute{IFISUR, Universidad Nacional del Sur, CONICET, Departamento de F\'isica - UNS, Av. L. N. Alem 1253, B8000CPB - Bah\'ia Blanca, Argentina. \and
IAFE, (UBA - CONICET), C1428EGA Buenos Aires, Argentina.}

\date{}

\abstract{In this work we present a theoretical study of the photoionization for atomic hydrogen confined in onion fullerene compared with the bare H atom and the single fullerene case. We obtained the expected confinement resonances for the integrated energy spectrum, finding different trends for the main peak and the first ATI peak integrations. We perform this calculations with a recently developed methodology using Generalized Sturmian Functions to numerically solve the time-dependent Schr\"odinger equation. 
} 

\maketitle

\section{Introduction}

In recent years, the study of the response to an electromagnetic field of an atom encapsulated in a fullerene cage has been the subject of several theoretical works (see for examples \cite{Dolmatov1,Dolmatov2}). Examples of situations and
phenomena of direct relevance to confined atoms are
helium droplets \cite{Zicovich-Wilson1994}, nano-size
bubbles formed in liquid helium \cite{Goodfriend1990},
atoms A encapsulated in hollow cages of carbon based
nano-materials, such as endohedral fullerenes A@C$_{60}$
\cite{Shinohara2000,Moriarty2001,Forro2001}, the chemical reactivity and chemical valence of
atoms \cite{Lawrence1981}, the reversible storage of
ions in certain solids \cite{Connerade1997,Connerade2000}, the appearance of helium bubbles
under high pressure in the walls of nuclear reactors
\cite{Walsh2000}, and ESR, NMR and magnetic
moments of compressed atoms \cite{Buchachenko2001},
etc. Confined atoms are also very close in principle
to the concepts underpinning confinement within
‘‘quantum dots’’ \cite{Johnson1995,Sako2003a,Sako2003}. Thus, the concept of a confined atom
provides insight into various problems of interdisciplinary significance.

A special case of the fullerene is that of the so-called fullerene onions or buckyonions, which consists of hollow carbon cages in which smaller buckyballs are encapsulated into larger ones (Cn@Cm@…). \cite{Xu2008,Echegoyen2010,Bartelmess2014}. Interestingly, fullerene onions have been detected in the interstellar medium \cite{DeHeer1993,Iglesias-Groth2004}, and were produced by synthesis by Ugarte in 1992 \cite{Ugarte1992}. Although confinement resonances in photoionization spectra of A@C$_60$’s have been studied extensively, little is known about confinement resonances in photoionization spectra of atoms confined in fullerene onions.

The model-potential method was a first step to study and qualitatively predict the confining effects of the cage on the spectral and dynamic properties of the atom. The external environment imposed by the fullerene cage can, in many instances, be described quite well by a simple, local, spherically symmetric, attractive cage potential that is generally taken to be of
constant depth in the region of the fullerene. Photoionization of atoms which are influenced by this model potentials has been treated using a number of methodologies, such as Hartree-Fock (HF) \cite{Dolmatov1,Dolmatov4}, random-phase approximation (RPA) \cite{Dolmatov1,Amusia1,Amusia2}, time-dependent close-coupling (TDCC) \cite{Ludlow1,Lee1}, R-matrix \cite{Gorczyca2012}, and matrix iterative method \cite{Grum2011}.

In a previous work \cite{frapiccini2017}, we proposed to use the Generalized Sturmian Functions (GSF) \cite{RevStur} to numerically solve the time-dependent Schrödinger equation of a caged atom interacting with a laser pulse. The adaptability of the GSF is key in this methodology, since we can use exponential decaying (real) Sturmians to solve the time-dependent portion of the problem, and the outgoing wave (complex) Sturmians for the time-independent term. To propagate the time-dependent wave packet during the interaction with the pulse, we use and explicit integrating scheme known as Arnoldi \cite{Saad1992,Frapiccini2014}, which is a Krylov subspace method. The methodology presented proved to be a useful approach to solve the TDSE avoiding the use of integrals to
obtain ionization amplitudes. 

Following this work, we propose to use the GSF methodology to study the photoionization of atomic H encapsulated in fullerene onions. We compare the photoionization spectra for the H@C$_{60}$ and H@C$_{240}$ with the onion structure H@C$_{60}$@C$_{240}$, analyzing the confinement resonances for the main peak and the first ATI peak separately. 

In Sec.2 we present an outline of the methodology to
extract the ionization amplitude and the spectral density
for a one-electron atomic system. In Sec. 3 we present results  for the study of avoided crossings due to
the change in the depth of the spherical well model potential
used to represent the effect of the fullerene. To our knowledge, no research on the avoided crossing due to two spherical wells has been done before. 
We also present results for the photoinization of the caged H, and concentrate the in the analysis of the confinement resonances in the photoionization spectra, calculating separately the contribution for the main (L=1) peak and first ATI (L=0) peak. 

Atomic units are used throughout unless otherwise indicated.

\section{Theory} 

\subsection{Time-independent Schr\"odinger equation: bound states}\label{sec1}

The effect of the confinement of the hydrogen atom by a fullerene shell is
modeled by a spherical well
\begin{equation}
V_{w}^{(i)}(r)=
\begin{cases}
-U_{0}^{(i)} \qquad \text{if} \qquad r_{c}^{(i)}\leq r \leq r_{c}^{(i)}+\Delta^{(i)} \\
0 \qquad \text{otherwise}
\end{cases}
\label{eq27}
\end{equation}
where $r_{c}^{(i)}$ is the inner radius of the well and $\Delta^{(i)}$ its thickness. In the case of the onion shells,
the potential can be written as
\begin{equation}
V_{w}(r)=\sum_{i=1}^{n}V_{w}^{(i)}(r)
\end{equation}
for an $n$-walled fullerene cage. For our study we will restrict the potential to $n=2$.

The bound states of the caged atom are obtained by solving the time-independent Schr\"odinger equation (TISE)
\begin{equation}
\left[H_{0}+V_{w}\right]\Psi_{\nu}(\mathbf{r})=E_{\nu}\Psi_{\nu}(\mathbf{r}) \label{eq28}
\end{equation}
with
\begin{equation}
H_{0}=T-\frac{Z}{r} \label{eq5}
\end{equation}
where $T$ is the operator of the kinetic energy and $Z$ is the atomic charge.

To solve the TISE Eq.~(\ref{eq28}) we expand the wave function in spherical coordinates, and use generalized
Sturmian Functions (GSF) \cite{Frapiccini2007,RevStur}
\begin{equation}
\Psi_{\nu}(\mathbf{r})=\sum_{jlm}a_{jl}^{\nu}\frac{S_{jl}(r)}{r} Y_{l}^{m}(\widehat{\mathbf{r}}) \label{eq29}.
\end{equation}
The generalized eigenvalue problem to solve is
\begin{equation}
\left[\mathbf{H_{0}}+\mathbf{V_{w}}\right]\mathbf{a}_{l}^{\nu}=E_{\nu,l}\mathbf{B}\mathbf{a}_{l}^{\nu} \label{eq29a}
\end{equation}
for each angular momemtum $l$ and $m=0$. All matrices are real and symmetric, and the overlap term is also positive definite.
The size of the matrices depend on the number $N_{max}$ of GSF in the expansion (\ref{eq29}), so the dimension of the matrices
will be $N_{max} \times N_{max}$.

\subsection{Time-dependent Schr\"odinger equation: photoionization} \label{sec2}

We write the time-dependent Schr\"odinger equation (TDSE) for a two particle system interacting with an external field in the form
\begin{equation}
\imath \frac{\partial}{\partial t}\Psi(\mathbf{r},t)=H(\mathbf{r},t)\Psi(\mathbf{r},t) \label{eqTDSE}
\end{equation}
where the Hamiltonian can be written as
\begin{equation}
H(\mathbf{r},t)=H_{0}+V_{w}+H_{int} \label{eqHamilt}
\end{equation}
with $H_{0}+V_{w}$ the unperturbed Hamiltonian, and $H_{int}$ the interaction with the field. The interaction
with the field of an electromagnetic pulse of finite duration is
\begin{equation}
H_{int}=
\begin{cases}
f(\mathbf{r},t) \quad \text{for} \quad t_{0}\le t \le t_{final} \\
0 \quad \text{for} \quad t> t_{final} \label{eqField}
\end{cases}
\end{equation}
with $\mathbf{r}$ being the electronic coordinates. 

The interaction with the pulse, within the dipole approximation, is written using the velocity gauge, and we consider here linear polarization
in the $\widehat{\mathbf{z}}$ axis, thus
\begin{equation}
H_{int}(\mathbf{r},t)= -\imath \mathbf{A}(t) \cdot \nabla_{\mathbf{r}}=-\imath \lvert \mathbf{A}(t) \rvert \frac{\partial}{\partial z} \label{eqgauge}
\end{equation}
For a photon energy $\omega$ and a pulse of duration $\tau$ we write
\begin{equation}
\lvert \mathbf{A}(t) \rvert= A_{0} g(\omega,t) \sin(\omega t) \quad \text{for} \quad t\in[0,\tau] \label{eqpulse}.
\end{equation}

To solve the TDSE Eq.~(\ref{eqTDSE}) for $t<t_{final}$ we expand the wave packet in spherical coordinates, and use 
GSF
\begin{equation}
\Psi(\mathbf{r},t)=\sum_{nl} a_{nl}(t)\frac{S_{nl}(r)}{r} Y_{l}^{0}(\widehat{\mathbf{r}}) \label{eqexp}
\end{equation}
with $a_{nl}(t)$ the expansion coefficients that depend on time and $Y_{l}^{m}$ the spherical harmonics.
The set of GSF used to solve the TDSE are real and with exponential decaying behavior at large distances. Details of the methodology involved in solving Eq.~(\ref{eqTDSE}) with the GSF can be found in a previous article \cite{frapiccini2017}.

The differential probability for an electron having the energy $E$ is determined in terms of the spectral density $D(E,t)$:
\begin{equation}
dP = D(E,t) dE, \label{eqdifprob}
\end{equation}
where
\begin{equation}
D(E,t) = \sqrt{2 E}\int \lvert C(\mathbf{k})\rvert^{2} d\Omega_{k} \label{eqDE}
\end{equation}
with $\Omega_{k}$ denoting the solid angle under which the electron is emitted and the coefficient $C(\mathbf{k})$ the ionization amplitude.

To obtain the coefficients $C(\mathbf{k})$, we find the evolution of the wave packet at $t>t_{final}$, which  is equivalent to solve the time-independent Schr\"odinger equation given by
\begin{equation}
(E-H_{0}-V_{w}) \Psi_{sc}(\mathbf{r})=\Psi(\mathbf{r},t_{final}) \label{eqscat}
\end{equation}
where $H_{0}+V_{w}$ is the atomic (time-independent) Hamiltonian, $\Psi_{sc}$ is a scattering term with outgoing boundary conditions, and
$\Psi(\mathbf{r},t_{final})$ is the wave packet
at the end of the pulse \cite{McCurdy2007}. 

To see how to extract the coefficients $C(\mathbf{k})$ from Eq.~(\ref{eqscat}), we write this equation by means of the Green's function
\begin{equation}
\Psi_{sc}(\mathbf{r})=\frac{1}{(E-H_{0})}\Psi(\mathbf{r'},t_{final})=G^{+}(\mathbf{r},\mathbf{r}') \Psi(\mathbf{r'},t_{final}) \label{eq9}
\end{equation}
Using the properties of the Coulomb Green's function, we can see that the asymptotic form of the scattering function is
\begin{equation}
\Psi_{sc}(\mathbf{r})\xrightarrow{r\rightarrow \infty} -\sqrt{2 \pi} C(k\widehat{\mathbf{r}})\frac{e^{\imath[kr+(Z/k)\ln 2kr]}}{r} \label{eq10}
\end{equation}
for an electron ejected with momentum $k=\sqrt{2E}$. This means that if the scattering function has the correct (outgoing wave) asymptotic
behavior, the ionization amplitude can be extracted from the function at sufficiently large values of the radius $r$.

\section{Results}

\subsection{Avoided crossing}

For this study, we fix the values of $r_{c}^{(1,2)}$ and $\Delta^{(1,2)}$ and calculate the energy eigenvalues in Eq.(\ref{eq29a}) for $U_{0}^{(1,2)}$ in a range from $0$ to $2$ a.u.. 
We use the data provided by Xu et al \cite{Xu1996} for a fullerene molecule C$_{60}$, which is $r_{c}^{(1)}=5.75$ a.u. and $\Delta^{(1)}=1.89$ a.u.. The second, outer
fullerene is located in place of the C$_{240}$ molecule, with values $r_{c}^{(2)}=12.6$ a.u. and $\Delta^{(2)}=1.9$ a.u. \cite{Dolmatov4}.

\begin{figure}[ht]
\begin{center}
\includegraphics[width=8.5cm,angle=0]{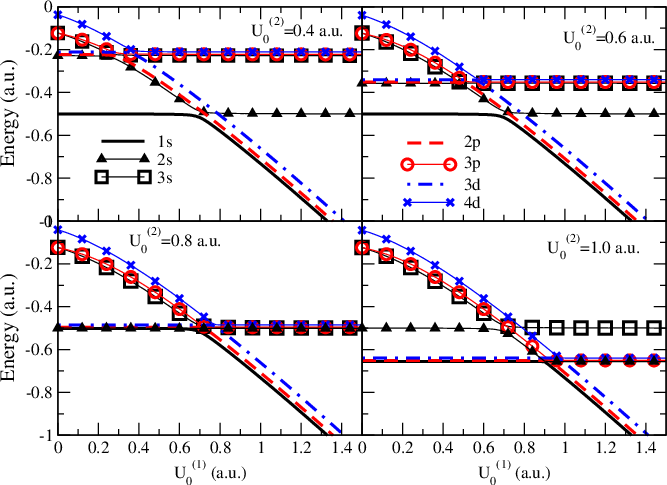}
\end{center}
\caption{(Color online) Energies of the H atom confined in two spherical well with $r_{c}^{(1)}=5.75$ a.u., $\Delta^{(1)}=1.89$ a.u., 
$r_{c}^{(2)}=12.6$ a.u. and $\Delta^{(2)}=1.9$ a.u., as a function of $U_{0}^{(1)}$ for four different values of $U_{0}^{(2)}$.} 
\label{fig1}
\end{figure}

In Fig.~\ref{fig1} we show the results for the firsts  $s,p$ and $d$ bound state energy levels. We fixed the value of the second well $U_{0}^{(2)}$
and plotted the energy as a function of the first well $U_{0}^{(1)}$. We can see here how the energies behave as the depth of the second well is increased.
The first we can notice is that, unlike in the case of a single fullerene cage, we have avoided crossings in the $p$ and $d$ levels. We also observe
that the $1s-2s$ crossing remains unchanged until the depth of the second well reaches a value of $U_{0}^{(2)}\approx 0.8$ a.u., and then
this crossing starts to 'move' to higher values of $U_{0}^{(1)}$.

\begin{figure}[ht]
\begin{center}
\includegraphics[width=6.0cm,angle=0]{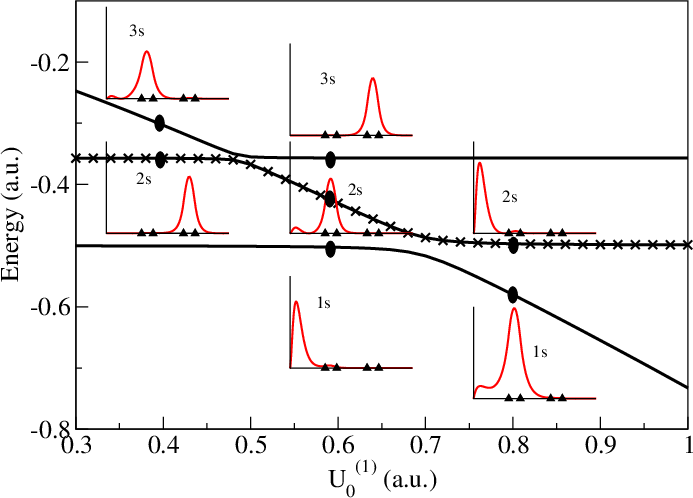}
\includegraphics[width=6.0cm,angle=0]{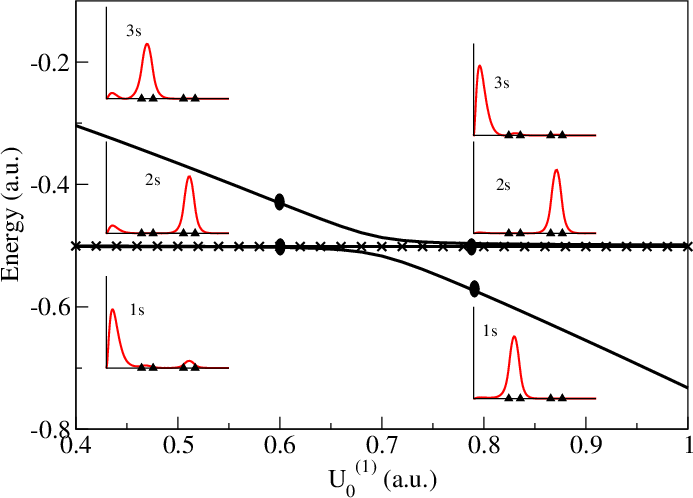}
\includegraphics[width=6.0cm,angle=0]{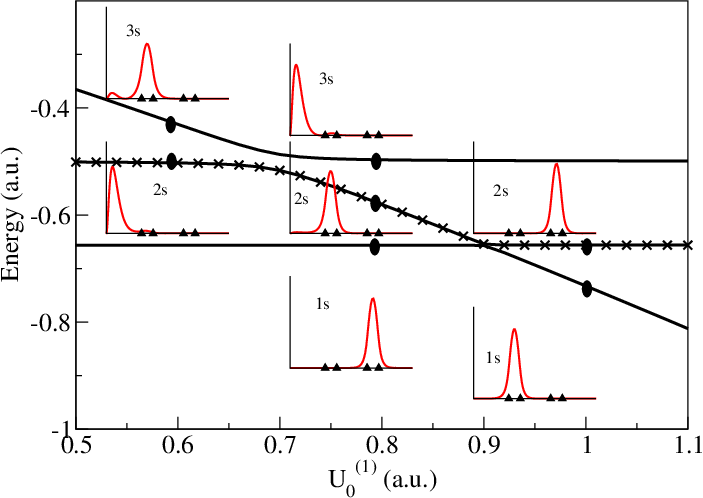}
\end{center}
\caption{(Color online) Radial $s-$state probability of the H atom confined in two spherical well with $r_{c}^{(1)}=5.75$ a.u., $\Delta^{(1)}=1.89$ a.u., 
$r_{c}^{(2)}=12.6$ a.u. and $\Delta^{(2)}=1.9$ a.u. near the energy crossing for $U_{0}^{(2)}=0.6$ (top), $U_{0}^{(2)}=0.8$ (center) and $U_{0}^{(2)}=1.0$ (bottom).} 
\label{fig2}
\end{figure}

In Fig.~\ref{fig2} we can see the radial probability density $\int r^{2} \lvert \Psi_{\nu}(\mathbf{r})\rvert^2 d\Omega_{r}$ 
for the $1s$, $2s$ and $3s$ states in the vicinity of the avoided crossings. 
In the top panel in Fig.~\ref{fig2}, for $U_{0}^{(2)}=0.6$ a.u., there is a 'mirror collapse' between the $2s-3s$ wave functions
first and then between the $1s-2s$. As the depth of the second well is increased, for $U_{0}^{(2)}=0.8$ a.u. in the center panel in Fig.~\ref{fig2} ,
the 'mirror collapse' occurs only between the $1s-3s$ levels, while the $2s$ wave function remains unchanged. For $U_{0}^{(2)}=1.0$ a.u. in the bottom
panel in Fig.~\ref{fig2} we go back to the same ordering as before, $2s-3s$ crossing and then $1s-2s$ crossing.
We can also see how, for the first two plots (up to $u_{0}^{(2)}=0.8$), we always start with the $1s$ state in the Coulomb well, the $2s$ state
in the second fullerene well and the $3s$ state in the first fullerene well. For higher values of $U_{0}^{(2)}$, the $1s$ and $2s$ are exchanged and now then$1s$ 
starts in the second well while the $2s$ starts in the Coulomb well.

\begin{figure}[ht]
\begin{center}
\includegraphics[width=6.0cm,angle=0]{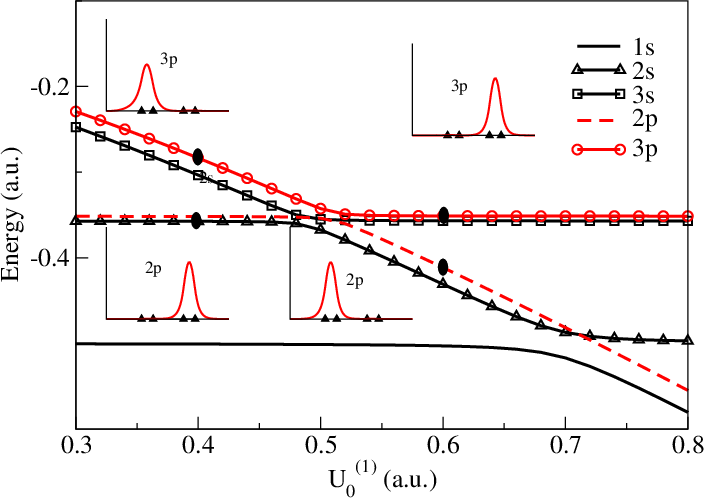}
\includegraphics[width=6.0cm,angle=0]{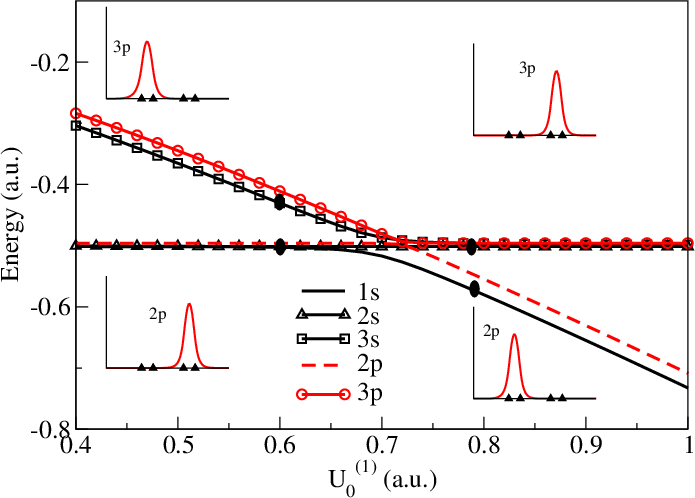}
\includegraphics[width=6.0cm,angle=0]{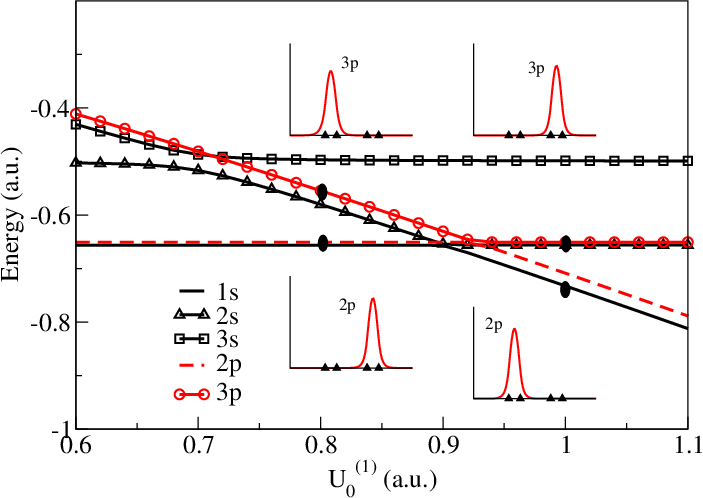}
\end{center}
\caption{(Color online) Radial $p-$state probability of the H atom confined in two spherical well with $r_{c}^{(1)}=5.75$ a.u., $\Delta^{(1)}=1.89$ a.u., 
$r_{c}^{(2)}=12.6$ a.u. and $\Delta^{(2)}=1.9$ a.u. near the energy crossing for $U_{0}^{(2)}=0.6$ (top), $U_{0}^{(2)}=0.8$ (center) and $U_{0}^{(2)}=1.0$ (bottom).} 
\label{fig3}
\end{figure}

In Fig.~\ref{fig3} we plotted the radial probability density for the $2p$ and $3p$ states near the avoided crossings for the same cases as the $s$ states.
We can see here how, for lower values of $U_{0}^{(1)}$, the $2p$ state is always located in the second fullerene well, 
while the $3p$ is in the first fullerene well. As the depth of the first well increases, we observe the mirror collapse between these bound states.
It is also observed how the energy crossing for the $2p-3p$ levels 'follows' first that of the $2s-3s$ levels, and after $U_{0}^{(2)}=0.8$ follows
that of the $1s-2s$ levels.

The avoided crossing phenomena is a mechanism for the state energy reordering, manifested by the energy level repulsion:
neighboring energy levels with the same symmetry do not cross each other, but rather come close and repel each other in an avoided crossing. 
An additional indicator of the external effects resides in the informational character. 
In information theory, entropy is a measure of the uncertainty
associated with a random variable. 

In this field, the term usually
refers to the Shannon entropy, which measures the expected value
of the information contained in a message, usually in units such as
bits, i.e., it is a measure of the average information content that is
missing when the value of the random variable is unknown.

The Shannon information entropy of one-normalized electron density in the coordinate space \cite{Shannon1948} is defined as
\begin{equation}
 S_{r}=\int \rho(\mathbf{r}) \ln\left[\rho(\mathbf{r})\right] d\mathbf{r} \label{eqShannon}
\end{equation}
where the electron atomic density is defined as
\begin{equation}
\rho(\mathbf{r})=\left\lvert\Psi_{\nu}(\mathbf{r}) \right\rvert^{2} \label{eqDens}
\end{equation}

This quantity is an information measure of the spatial delocalization 
of the electronic cloud. So, it gives the uncertainty of the localization of the electron. 
The lower this quantity, the more concentrated the wave function of the state, 
the smaller the uncertainty, and the higher the accuracy in predicting the localization of the electron. 

The variation of the Shannon entropy of states with an external potential strength may lead to gaining a deeper physical
insight into the dynamics of the system through the avoided crossing region \cite{PhysRevLett.91.113001}.

\begin{figure}[ht]
\begin{center}
\includegraphics[width=6.0cm,angle=0]{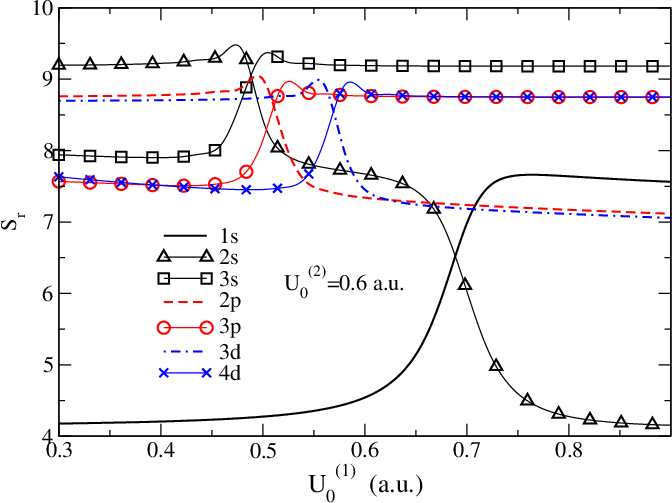}
\includegraphics[width=6.0cm,angle=0]{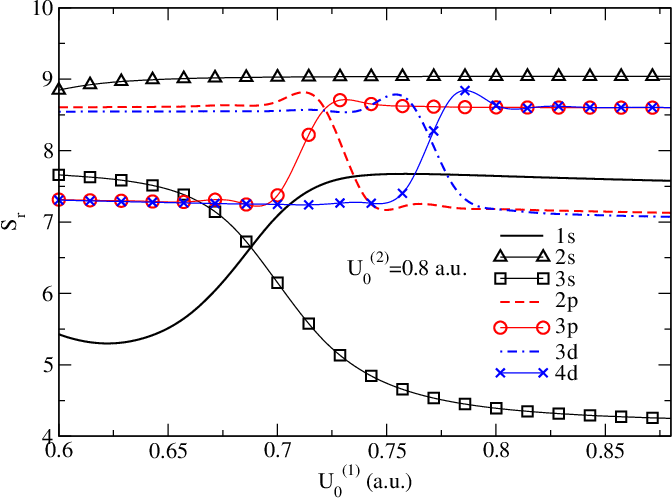}
\end{center}
\caption{(Color online) Shannon entropy of the H atom confined in two spherical well with $r_{c}^{(1)}=5.75$ a.u., $\Delta^{(1)}=1.89$ a.u., 
$r_{c}^{(2)}=12.6$ a.u. and $\Delta^{(2)}=1.9$ a.u. for $U_{0}^{(2)}=0.6$ (top) and $U_{0}^{(2)}=0.8$ (bottom).} 
\label{fig4}
\end{figure}

In Fig.~\ref{fig4} is shown the Shannon information entropy as a function of the first fullerene cage strength for two different values
of the outer fullerene cage depth.
The same energy levels as in Fig.~\ref{fig1} are plotted.
We can see here how the states exchange their informational properties as they go through an avoided crossing.

\subsection{Confinement resonances in photoionization}

We now turn to the study of the photoionization of the confined atoms, 
using the methodology described in Sec.\ref{sec2}. The interaction with the pulse is as described in Eq.~\ref{eqpulse} with a sine
square envelope $g(\omega,t)=\sin\left({\frac{\pi}{\tau}t}\right)^2$ with a total duration of $\tau=2\pi n_{c}/\omega$, where n$_c$ is an integer 
giving the number of optical cycles.

We consider photoionization from initial state $1s$. The values for $r_{c}^{(1)},r_{c}^{(2)},\Delta^{(1)}$ and
$\Delta^{(2)}$ are the same as in the previous section, and for the the C$_{60}$ cage, we use the parameters of the model potential found in \cite{Xu1996,0953-4075-38-7-003}: 
$U_{0}=8.22$ eV, and for the C$_{240}$ \cite{Dolmatov4}: $U_{0}=10$ eV.

In all the calculations presented now the peak intensity $5\times 10^{14}$ W/cm$^2$ and $n_{c}=16$ optical cycles. All the calculations shown 
in this section were performed using $N_{max}=300$ and $l_{max}=10$, and the size of the Krylov space was $n_{Krylov}=30$, with a fixed time
step of $\delta t=0.03$. 

\begin{figure}[ht]
\begin{center}
\includegraphics[width=6.0cm,angle=0]{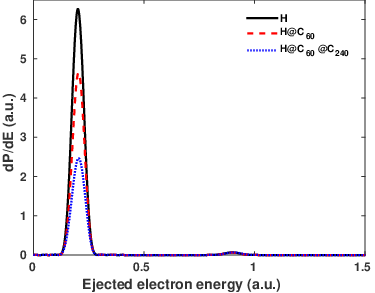}
\includegraphics[width=6.0cm,angle=0]{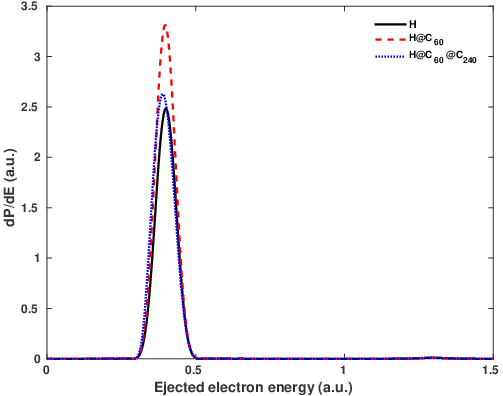}
\end{center}
\caption{(Color online) Energy spectrum of the ejected electron for photon energies of $\omega=0.7$ (top) and $0.9$ (bottom) a.u., for the confined and bare $H$ atoms. } 
\label{fig5}
\end{figure}

\begin{figure}[ht]
\begin{center}
\includegraphics[width=9.0cm,angle=0]{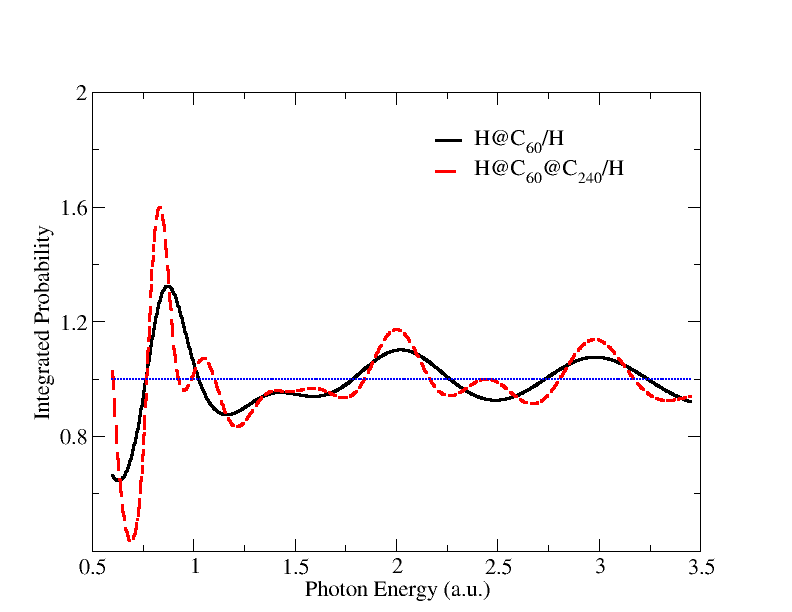}
\end{center}
\caption{(Color online) Integrated energy spectrum for the main peak as a function of the photon energy. } 
\label{fig6}
\end{figure}

As a first test we performed calculations for photon energies of $\omega=0.7$ and $0.9$ a.u.. The endohedral atoms H@C$_{60}$ and H@C$_{60}$@C$_{240}$ were compared to the bare H atom. 

The results for the ejected electron energy spectrum are shown in Fig.~\ref{fig5}. It is observed here how the intensity of the line is affected by the presence of the cage(s). In the case of $\omega=0.7$ a.u. for example, we see that the endohedral atom H@C$_{60}$ has a lower intensity than that of the H atom, as well as the H@C$_{60}$@C$_{240}$ . On the other hand, for $\omega=0.9$, both intensities are higher than that of the H. 

\begin{figure}[ht]
\begin{center}
\includegraphics[width=7.0cm,angle=0]{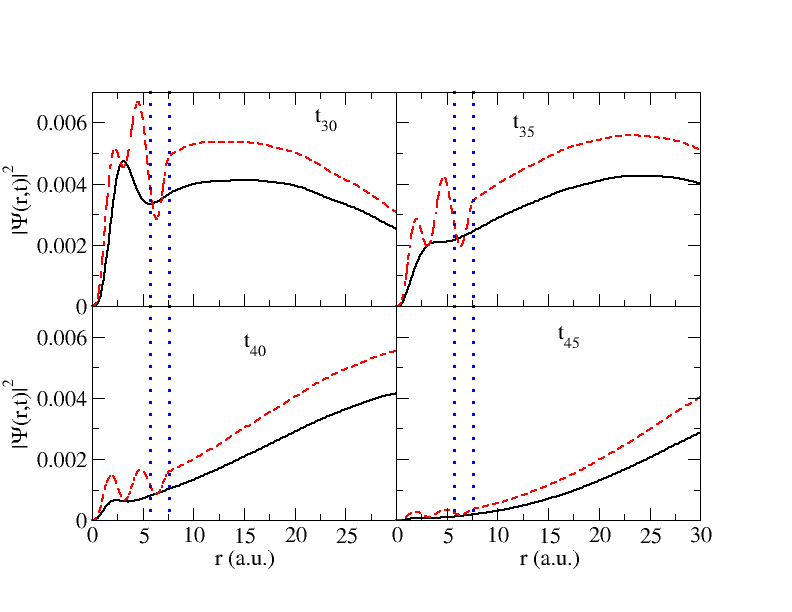}
\includegraphics[width=7.0cm,angle=0]{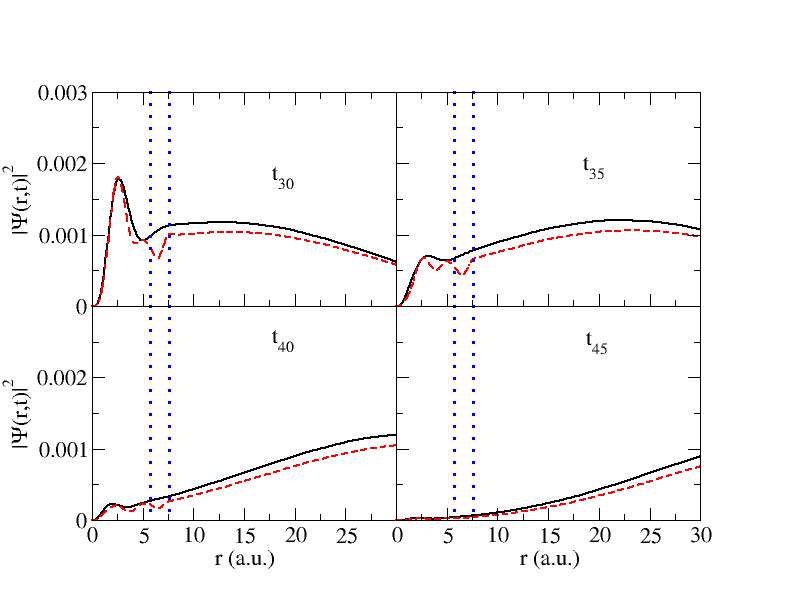}
\end{center}
\caption{(Color online) Radial probability density for the L=1 wave
for the bare atom H and endohedral H@C$_{60}$ near a maximum resonance (top)
and minimum resonance (bottom), for four different time stages during the propagation of the wave packet. Dotted (blue) line shows the location for the C$_{60}$ cage.} 
\label{fig7}
\end{figure}

This oscillations in intensity of the spectrum as a function of the photon
(photoelectron) energy is due to the \textquoteleft confinement resonances\textquoteright.
As discussed by Connerade et al \cite{0953-4075-33-12-309}, they studied the origin and properties of confinement resonances considering hydrogenic ions placed at the centre of a spherical shell as in Eq.~(\ref{eq27}) for a single well. They assume that the resonances are due to features
in final (continuous) electronic states owing to the confinement. 

In Fig.(\ref{fig6}) we show the results for the integration of the energy spectrum $dP/dE$ around the main peak as a function of the photon energy, for the  H@C$_{60}$  and  H@C$_{60}$@C$_{240}$. We can see here how the presence of the cage(s) affects the intensity of the main peak. The range of photon energy in Fig.(\ref{fig6}) corresponds to electrons emitted with energies lower than 3 a.u. in order to ensure that the electron wavelength is
larger than the distance between the carbon cage atoms \cite{Amusia_2008}. The addition of the second well for the H@C$_{60}$@C$_{240}$ increases the number of resonances, as expected. 

In Fig.(\ref{fig7}) we see the time evolution of the L=1 wave of the radial
probability density for the H atom and the encapsulated H@C$_{60}$. The top panel in Fig.(\ref{fig7}) corresponds to a photon energy of $\omega=0.87$ a.u. near a maximum in Fig. (\ref{fig6}) and the bottom panel to $\omega=1.16$ a.u. near a minimum in Fig.(\ref{fig6}). It is clear here the effect of the cage in the constructive (top panel) and destructive (bottom panel) interference in the inner region $0<r<r_{c}^{1}$. 

\begin{figure}[ht]
\begin{center}
\includegraphics[width=7.0cm,angle=0]{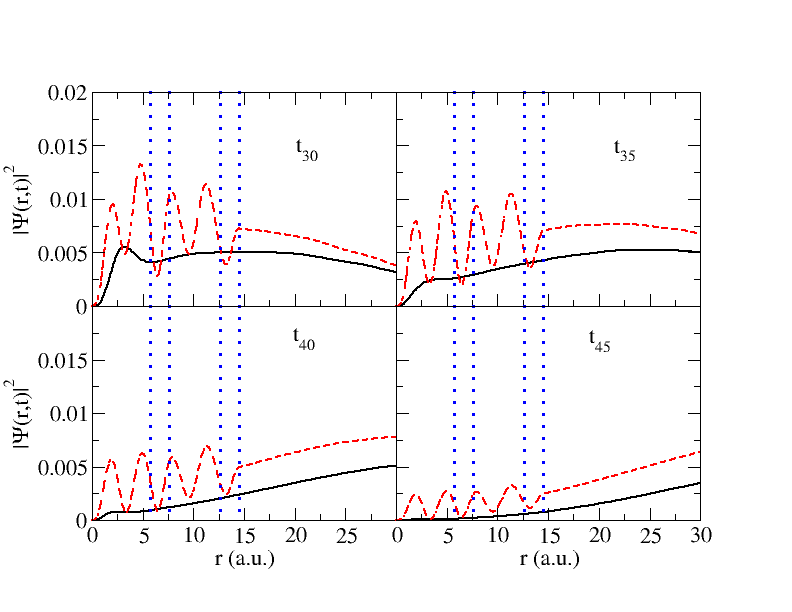}
\includegraphics[width=7.0cm,angle=0]{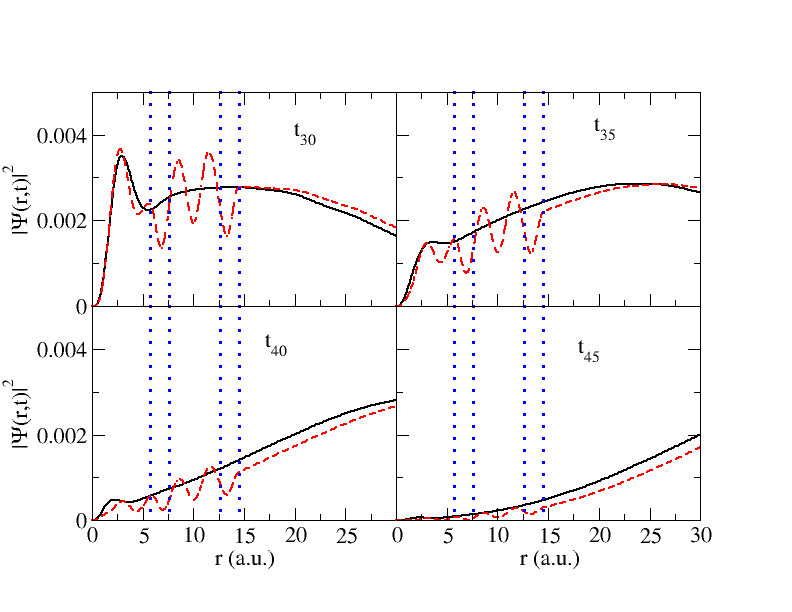}
\end{center}
\caption{(Color online) Radial probability density for the L=1 wave
for the bare atom H and endohedral H@C$_{60}$@C$_{240}$ near a maximum resonance (top) and minimum resonance (bottom), for four different time stages during the propagation of the wave packet. Dotted (blue) line shows the location for the C$_{60}$ and C$_{240}$ cages.} 
\label{fig8}
\end{figure}

In Fig.(\ref{fig8}) we see the time evolution of the L=1 wave of the radial
probability density for the H atom and the encapsulated H@C$_{60}$@C$_{240}$. The top panel in Fig.(\ref{fig8}) corresponds to a photon energy of $\omega=0.83$ a.u. near a maximum in Fig. (\ref{fig6}) and the bottom panel to $\omega=0.96$ a.u. near a minimum in Fig.(\ref{fig6}). Here we see the effect of the cage in the constructive (top panel) and destructive (bottom panel) interference in the inner region $0<r<r_{c}^{1}$ and also in the region between the two cages. 

\begin{figure}[ht]
\begin{center}
\includegraphics[width=7.0cm,angle=0]{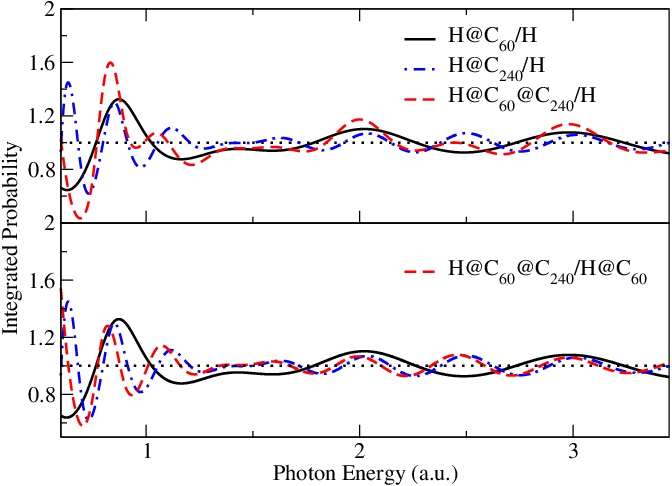}
\end{center}
\caption{(Color online)Integrated energy spectrum for the main peak (P peak) as a function of the photon energy. Top panel shows the ratio with respect to the bare H atom, bottom panel shows the single walled H@C$_{60}$ and H@$_{240}$ with respect to H, and the double walled H@C$_{60}$@C$_{240}$ with respect to H@C$_{60}$.} 
\label{fig9}
\end{figure}

In \cite{Dolmatov4}, the study of multiwalled fullerenes suggested a trend in the confinement oscillations as the atom was encapsulated in larger fullerenes. In Fig. (\ref{fig9}) we see how we can draw the same conclusions as in \cite{Dolmatov4} for the integrated main line (P peak), comparing the three cases of H@C$_{60}$@C$_{240}$/H with the single walled H@C$_{60}$/H and H@C$_{240}$/H. For the caged H atom,  in the top panel of Fig. (\ref{fig9}) we see how the single walled atoms oscillate around the dotted line (equal to 1), but the H@C$_{60}$@C$_{240}$ actually seems to oscillate on the H@C$_{60}$ line. To test this, in the bottom panel of Fig.(\ref{fig9}) we plotted H@C$_{60}$@C$_{240}$/H@C$_{60}$ (red broken line), and it clearly shows a behavior similar to the H@C$_{240}$/H, except for lower photon energies. 

\begin{figure}[ht]
\begin{center}
\includegraphics[width=7.0cm,angle=0]{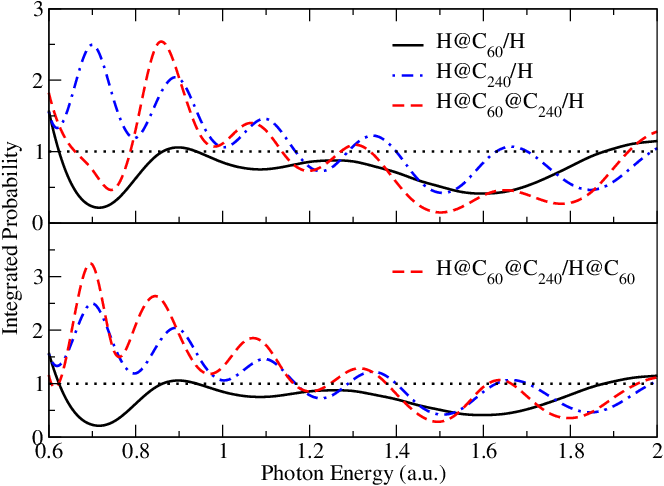}
\end{center}
\caption{(Color online) (Color online) Integrated energy spectrum for the first ATI peak (S peak) as a function of the photon energy. Top panel shows the ratio with respect to the bare H atom, bottom panel shows the single walled H@C$_{60}$ and H@$_{240}$ with respect to H, and the double walled H@C$_{60}$@C$_{240}$ with respect to H@C$_{60}$.} 
\label{fig10}
\end{figure}

To see if the trend observed for the integrated main line extends to the first ATI peak (S peak), we calculated the same ratios as in Fig. (\ref{fig9}) for this peak. 
If Fig.(\ref{fig10}) we can see the same behavior for the ATI peak as for the main line, meaning that the ratio of H@C$_{60}$@C$_{240}$/H@C$_{60}$ behaves as the H@C$_{240}$/H for high photon energies. We see here that the behavior of the P line is not the same as the S line, the location and number of peaks differ. This is mainly due to the intermediate states involved in the two-photon process, which is the primary source of the S peak.

\section{Conclusions}

We apply an \textit{ab-initio} methodology to solve the time-dependent Schr\"odinger equation of an atom interacting with an electromagnetic pulse of finite duration. The approach is based on the Generalized Sturmian Functions \cite{frapiccini2017}, and their adaptability to define different asymptotic behaviors. 

We present an application by studying the influence of the confinement
of the H atom in a fullerene cage C$_{60}$ and C$_{240}$ compared with the fullerene onion C$_{60}$@C$_{240}$. First we perform a study of the avoided crossing for the onion
with a simple spherical well potential to represent the fullerenes. Since we have now two parameters to vary (the depth of each well), we fix the depth of in the location of the C$_{240}$ cage at several different values, and plot the energy eigenvalues as a function of the depth of the well of the C$_{60}$ . For the cases shown, we observe crossings not only for the $s$ states, as was the case for a single cage, but also crossings in the $p$ and $d$ states. We also show calculations for the Shannon information entropy for the electron density near some of the crossings, which clearly display the exchange of the informational properties as they go through the avoided crossing.

Finally, we calculate the ionization of the bare H and caged atom interacting with an electromagnetic pulse in the range of 0.6 a.u. to 3.4 a.u. of photon energy. We present the results for the energy spectrum integrated around the main peak as a function of the photon energy for the H@C$_{60}$ and H@C$_{60}$@C$_{240}$ with respect of the bare H atom. We obtained the expected confinement resonances, observing the increase in the number of oscillations in the onion fullerene with respect to the single well fullerene. The plots of the radial probability for the L=1 wave in the vicinity of a maximum of minimum in the energy spectrum show the constructive and destructive interference which give origin to the resonances. 
The results for the main peak show clearly that at high photon energies, we can separate the effect of the inner C$_{60}$ and outer C$_{240}$ cages for the onion fullerene, which is not the case for lower photon energies. 

We perform the same calculations for the first ATI peak, observing the confinement resonances, but at different locations. However, we confirm the same behavior for high photon energies, in which the outer fullerene and inner fullerene effects can be separated. 

The calculations performed in encapsulated hydrogen are a useful starting point to study the influence of the onion fullerenes in other atomic systems. The conclusions drawn here are applicable to other atomic systems whose initial state has a similar probability distribution to that of the 1s. While our focus here is to present results for the confined H, other multielectronic atoms could be represented by means of the one-active electron approximation.  

\section*{Acknowledgments}

We acknowledge the support by Grant No. PIP 201301/607
CONICET (Argentina) also thank the support by Grant
No. PGI (24/F059) of the Universidad Nacional del Sur.

\bibliographystyle{epj}
\bibliography{ReferencesAF}

\end{document}